# Improving Software Development Processes with Multicriteria Methods

Elena Kornyshova, Rébecca Deneckère, and Camille Salinesi

*\* Centre de recherche en Informatique, Université de Paris I Panthéon Sorbonne
90, rue de Tolbiac, 75013 Paris, France
kornyshova@univ-paris1.fr, denecker@univ-paris1.fr, salinesi@univ-paris1.fr*

RÉSUMÉ. *Tous les processus de développement de logiciels comportent des étapes incluant des choix, des prises de décisions. Il arrive que les méthodes utilisées offrent un certain guidage à l'ingénieur pour naviguer à travers ces choix. Cependant, de manière très courante, les arguments permettant de prendre la bonne décision sont extrêmement pauvres et le choix est finalement effectué de manière intuitive et hasardeuse. Le but de notre travail est d'offrir à l'ingénieur un guidage plus formel à l'aide de l'intégration et de l'application de méthodes multicritères dans le processus de développement de logiciels. Cette approche est illustrée par la sélection et l'application de priorisation aux risques, cas d'utilisation et outils dans le processus RUP.*

ABSTRACT. *All software development processes include steps where several alternatives induce a choice, a decision-making. Sometimes, methodologies offer a way to make decisions. However, in a lot of cases, the arguments to carry out the decision are very poor and the choice is made in an intuitive and hazardous way. The aim of our work is to offer a scientifically founded way to guide the engineer through tactical choices with the application of multicriteria methods in software development processes. This approach is illustrated with three cases: risks, use cases and tools within Rational Unified Process.*

MOTS-CLÉS : *Prise de décisions, Méthodes multicritères, Processus de développement des logiciels.*

**1. Introduction**

Researches on several engineering fields (systems engineering, process engineering, method engineering, and so on) show that there are many development cases where information system (IS) engineers has critical choices to carry out. As a matter of fact, they have to deal with a large number of characteristics, artifacts, ideas, possibilities, etc. Many strategies are offered to manage them and choosing one over the others is often a very difficult task to handle. Some development activities aim to sort possible alternatives by prioritizing them. However, these priorities are often applied intuitively and there is a great need for a better priorisation support.

Generally, a decision-making (DM) problem is defined by the presence of alternatives. The traditional approach consists in using only one criterion in order to select alternatives. The usual example is the selection of the projects according to the net present value. However, using a single criterion is not sufficient when the consequences of the alternatives to be analyzed are important (Roy, 1996). The goal of the Multicriteria (MC) DM methods consists in defining priorities between alternatives (actions, scenarios, projects) according to multiple criteria. In contrast to a monocriterion approach, MC methods allow a more in-depth analysis of the problem because they consider various aspects. However, their application has proved more difficult.

MC DM methods have shown their qualities for over 30 years (Berander, 2005) and they currently dominate in the field of decision-making (Baudry et al., 2002; Gomez_Limon et al., 2003). They appeared at the beginning of the Sixties, and their number and application contexts increase continually. For example, these methods are employed for requirements priorisation (Weigers, 1999), to choose evolution scenario (Papadacci et al., 2005), or to make operational decisions (Bouyssous, 2001).

Five families of MC methods can be considered: MAUT (Keeney et al., 1993), AHP (Saaty, 1980), outranking methods (Roy, 1996), weighting methods (Keeney, 1999), and fuzzy methods (Fuller et al., 1996). These methods will be detailed in the following.

We propose in this work to improve any development process with the use of multicriteria methods as a way to choose the most adapted alternative to each situation. We propose a process, illustrated by an example within Rational Unified Process (RUP) (Rational Rose, 2007; Kruchten, 1998), which integrates MC methods at the DM point of the development process. Our aim is to propose a formal approach for priorisation in order to enhance DM in development process.

The paper is organized as follows: section 2 gives an overview of our proposed process, which is illustrated in section 3 on three DM points of RUP, and concluded in section 4.

## 2. Overview of the Multicriteria Methods Integration Process

Our proposal consists of the integration of MC methods in the methodologies of software development. It is described by an "integration process" (IP) which is presented on Figure 1.

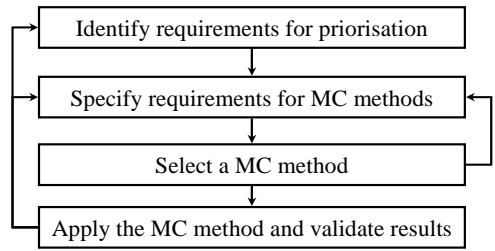

**Fig. 1.** *Process of integration of MC methods into software development methodologies.*

The integration process includes four steps: 1) Identify requirements for priorisation, 2) Specify requirements for MC methods, 3) Select a MC method, and 4) Apply the MC method and validate results. This IP includes both direct steps and flashbacks. The former indicate the normal IP development, and the latter enable returns to the previous steps if necessary.

### 2.1. Identify Requirements for Priorisation

This step may also be seen as the recognition and description of a specific situation of DM. The first element to define is the identification of the presence of alternatives. If a process offers a different manner to fulfill a specific objective, we may see this process as a "DM point". Identifying these points may be a difficult task to perform and we suggest asking the following questions:

- "What is the type of guidance to run this task: linear or tree form (set of possibilities)?"

- "Does the guidance offer arguments (metrics or criteria) to choose between the alternatives?"

- "Does the guidance offer a way to assign a prioritization to these alternatives?"

There are different kinds of DM problems. They may be classified (according to the number of criterion and of decision-makers they have) into five types (cf. Figure 2).

The first type presents a monocriterion problem and can be resolved as an optimization task. In the following, we will focus only on the problems that can be solved by MC methods (types: 2 to 5).

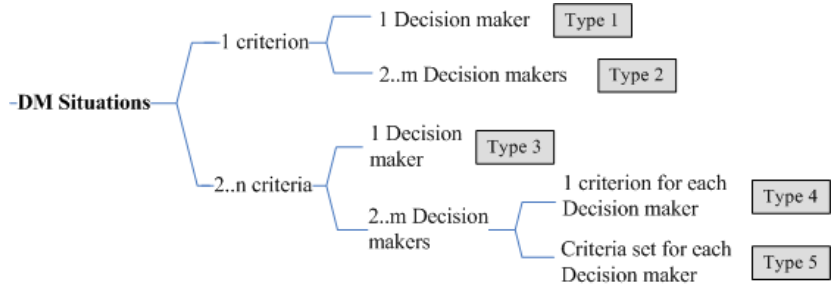

**Fig. 2.** *Typology of decision-making problems.*

When the DM point has been identified, the IP step guides the engineer in describing its situation. B. Roy defines three basic concepts that play a fundamental role in analysing and structuring decisions in close connection with the decision process itself (Roy, 2005): alternatives (potential actions), criteria family, and decision problem. Based on this, we propose to specify decision situation as a <Problem; Alternative; Criterion> triplet, where problem refers decision problem; alternative refers the collection of alternatives among which one will be chosen; and criterion refers the list of criteria by which alternatives will be evaluated. This description will allow the engineer to define the DM point on a generic level (called level 1 in this work).

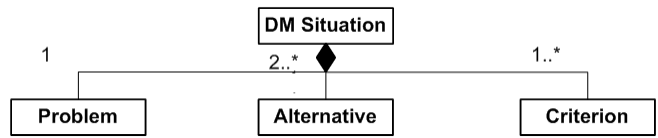

**Fig. 3.** *Model of DM situation.*

The decision *problem* (Roy, 2005) can be defined by the result expected from a DM. When the result consists in a subset of a potential alternatives (most often one alternative) then it is a *choice problem*. When the result represents the potential alternatives' affectation to some predefined clusters, then it is a *classification problem*. When the result consists in a potential alternatives ordered collection, then it is a *ranking problematic*. Given that each MC method is able to support a specific type of decision, it is important to know which type of decision is faced to be able to select the appropriate DM method. The concept of *alternative* designates the object of decisions. Any decision involves at least two alternatives that must be well identified. A *criterion* can be any type of information that enables the evaluation of alternatives and their comparison. Often, development processes already propose a predefined criteria set. This set can be improved by adapting it to the project at hand. One of the improvement possibilities takes its roots in two directions: software metrics (Mills, 2005) and typology of characteristics of IS development project (Kornyshova et al., 2007).

Within a MC problem, the metrics and the projects characteristics are considered as criteria. In a general way, the criteria may be qualitative or quantitative, relative or absolute, and criteria of time, cost, quality, size, efficiency, and so on.

**2.2. Specify Requirement for MC Methods**

In order to deal with decisions, we define a second level of decision-making for selecting a MC method (DM Situation L2). Whereas the level 1 deals with the prioritization problem, the level 2 is addressing the MC methods selection problem to solve the level 1 one. The identification of *requirement for MC methods* allows characterizing the specific parameters required for MC method selection. The problem is always a choice, the alternatives are MC methods, and the selection is made using criteria defined as (a) an aggregate view of the requirements for priorisation, and (b) supplementary criteria referring to the usage of the intended method. The Figure 4 illustrates the model of DM situation applied to the selection of MC method (L2 decision).

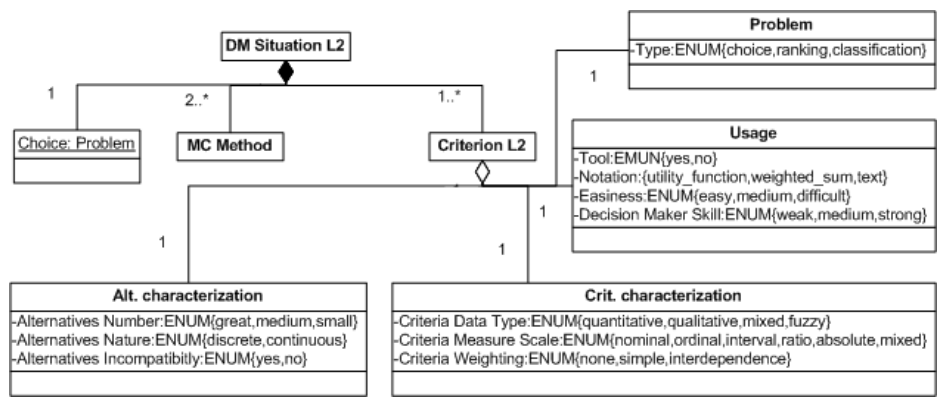

**Fig. 4.** *Model of DM situation for selecting MC method.*

Several strategies can be applied to specify requirements for MC methods. One of them is to specify the requirements *by problem investigation*. It means that the engineer has to identify the operations that enable to switch from the requirements for prioritization to the requirements for MC methods. These operations are (i) for problem: retaining the problem type; (ii) for alternatives: calculating the alternatives number, retaining alternatives nature, retaining alternatives incompatibility, and (iii) for criteria: retaining criteria data type, retaining criteria measure scale, and retaining weighting type. Additional information may also be required to specify the MC method usage in the given situation: if a DM tool is needed or not, the nature of the notation, the method easiness, and the level of engineer skills required for applying the MC method.

**2.3. Select a MC Method**

Each MC method is able to deal with problems with specific characteristics. For instance, the number and nature of the alternatives, the decision criteria or the presence of multiple stakeholders with different viewpoints. Besides, the existing methods have different characteristics such as complexity or ability to deal with quantitative or qualitative criteria. A few selection approaches were thus developed to guide specifically MCDM method selection. The state of the art is presented in (Kornyshova et al., 2008).

Our assumption is that a process guiding the selection of a DM method should (a) be simple to use, (b) provide results that can be trusted, and therefore (c) take into account all the relevant aspects of the situation at hand. Our approach focusing on these relevant aspects focuses on the comparison technique presented in the next sub-section. The current section focuses on the selection process itself.

We introduce the notion of MC method interface to guide MC method selection. The interface represents the characteristics of the situations in which a given MC method can be used and corresponds to the criteria set from the model presented in Fig. 4. The figure 5 shows the relationship between method and interface and several MC method family' interfaces, which are described in the Table 1. In this table, a line represents a general attribute of the interface (level 2) and a column represents a particular MC method family.

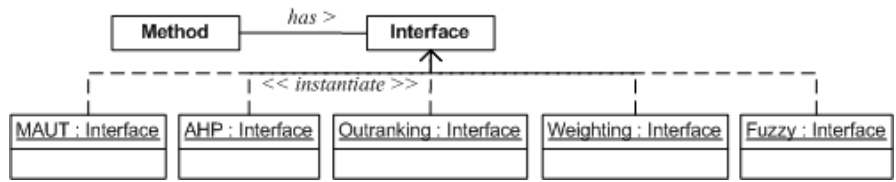

**Fig. 5.** *MC methods interfaces.*

Experience may be sufficient to select a method, in particular if the exact same situation has already been met.

An MC method may be selected by MC search. This means that the engineer has to search an appropriated method using L2 criteria identified earlier in order to obtain one or several MC methods corresponding to his/her requirements for MC method.

If the achievement of the MC search application drives to the selection of several MC methods, it is possible to choose one of them by weighting. Using this approach, weights must be given to the L2 criteria. These weights indicate the relative importance of the L2 criteria to the situation at hand. Then, "0" or "1" values are allocated to candidate MC methods according to each criterion. The method having the highest weighted sum of criteria values is then chosen. This strategy is not adequate when the previously selected methods have the same interfaces with reference to specified requirements.

**Table 1.** Instantiation of MC methods interfaces.

|  | MAUT | AHP | Outranking | Weighting | Fuzzy methods[1] |
|---|---|---|---|---|---|
| 1. "Problem" |  |  |  |  |  |
| 1.1. Choice | Yes | Yes | Yes | Yes | Yes |
| 1.2. Ranking | Yes | Yes | Yes | Yes | Yes |
| 1.3. Sorting | No | No | Yes | No | Yes |
| 2. "Potential actions" |  |  |  |  |  |
| 2.1. Number of alternatives | Great, medium, small | Small | Great, medium, small | Great, medium, small | Different |
| 2.2. Alternatives' set nature | discrete | discrete | discrete | discrete | Different |
| 2.3. Incompatibility and conflicts of alternatives | Yes | No | Yes | No | Different |
| 3. "Criteria" |  |  |  |  |  |
| 3.1. Data type | quant., qual. | quant., qual. | quant., qual. | quant. | Different |
| 3.2. Measure scale | Yes | No | Yes | No | Different |
| 3.3. Criteria weighting | Yes, simple | Yes, interdep | Yes, interdep | Yes, simple | Different |
| 4. "Usage" |  |  |  |  |  |
| 4.1. Tool | No | Yes | Yes | Yes | Different |
| 4.2. Notation | Utility function | Weighted sum | Textual | Weighted sum | Different |
| 4.3. Easiness of use | Difficult | Easy | Medium | Easy | Difficult |
| 4.4. Decision maker skills | strong | medium | strong | week | strong |

### 2.4. Apply the MC Method and Validate Results

The final step of our proposed process is to apply the chosen multicriteria methods on the identified decision points of the development process. The validation is made following the matching between the users' requirements and the obtained results. The MC methods application and its complexity degree depend on the selected method. It may require additional skills or the acquisition of a tool that supports MC decision making. The presence of a tool is an important factor for practitioners who are concerned with the rapid application of a selected MC method. Tools are however, sometimes costly (purchasing and training), and their acquisition and deployment can be time consuming.

The engineer may also execute the MC method by achieving manual calculation or by developing a tool ad hoc. Applying different methods involves different activities. For instance, the MAUT requires constructing partial utility functions and their aggregation into a general utility function by addition or multiplication (Keeney et al., 1993). AHP is based on a dominance hierarchy and carried out by decision-makers' pair-wise comparisons (Saaty, 1980). Outranking methods are based on analysis of the degree of dominance of one alternative over another (Bouyssous, 2001; Roy, 1996). Weighting methods are characterized by a weight assignment being applied to the decision criteria; and the aggregation of the evaluations is based on a weighted sum

---

[1] Fuzzy methods differ according to the "basic" MC method: MAUT, outranking methods, and so on. Hence, they have the value "Different".

(Keeney, 1999). The fuzzy MC methods employ the fuzzy sets theory to add flexibility and to enrich methods by fuzzy parameters (Fuller et al., 1996).

**3. Application Example with the Rational Unified Process**

We propose to illustrate the use of the proposed process by guiding decisions in the Rational Unified Process (RUP) (Rational Rose, 2007; Kruchten, 1998). The RUP is a body of software engineering practices, which is maintained on a regular basis to reflect changes in industry practices. It provides a wealth of guidance on software development practices that both novice and experienced practitioners can exploit. However, although many RUP practices call for decision-making, there is very little information about how to achieve these decisions. All these arguments, together with the fact that the RUP is widely used in the industry, convinced us that it was a good candidate to apply our approach and evaluate it. This paper presents details about the core elements of our proposal, which consists on identifying requirements for decision, specifying requirements for MC methods, and selecting MC methods.

Guidance is provided by the RUP under the form of descriptions of the tasks that can be achieved and of the best practices attached to them. Putting ourselves in the position of a person who wants to prepare a method for a project beforehand, we start by scanning each task described to find those offering alternatives and some kind of DM guidance. We chose to study 3 tasks more closely: (a) select and acquire tools, (b) prioritize use cases, and (c) analyze and prioritize risks[2].

**Select and Acquire Tools.** This task guides the adoption of tools that support other tasks in the RUP. Tools that need to be selected should fit the particular requirements of the organization for which the selection is made. Furthermore, special tools sometimes have to be developed internally to support special needs. One of the steps in this task is to collect information about tools in order to gain a better understanding. This information later serve as selection criteria to help the system engineer decide which tool is right for the project at hand. The criteria for tool selection are tool features, vendor and cost characteristics. The RUP proposes to grade each criterion for evaluating candidate tools. However, the guidance stops there and the engineer is left alone at the moment of the actual decision making.

**Analyze and Prioritize Risks.** This task describes how to identify, analyse and prioritize IS project risks. To achieve this, an inventory of what can go wrong within the project must be made. Events that might decrease the chance of delivering all the required IS features at the end of the project, at the required level of quality, and on time/within budget. The RUP guides this by telling how to (i) look within complementarities and redundancies to see if they would be a source of risk, (ii) put them in a table known as the Risk List, and (iii) rank risks in decreasing order of importance and associate them with specific mitigation or contingency actions. Again, the RUP is very vague with respect to this DM problem: an "order of importance" with respect to these criteria is not clearly defined.

---

[2] Our case study is nominative and simplified. It was elaborated specially for illustrating suggested approach application.

**Prioritize Use Cases.** The prioritization of use cases allows deciding their order of development. The RUP guidance proposes that the software architect selects a certain number of scenarios and use cases to be analyzed and designed. This proposal is completed and refined in several ways: by development teams, customer requirements, and based on COTS products. The selection is then made by characterizing key factors. For instance, architecturally significant use cases that are poorly understood or likely to change should be prioritized for clarification and stabilization.

These examples are presented in Table 2., which gives an overview of requirements for L1 decisions. Some considerations must be made. For instance, the cost evaluation of tools is carried out according to 5-grade scale (in RUP, - a 3-grade scale) for facilitating DM.

**Table 2. Examples description.**

| Task (task goal) | Criteria | Suggested method |
|---|---|---|
| Select Tools (select tools that fit the need of the project) | tool criteria (features and functions, integration, applicability, extendibility, team support, usability, quality, performance, maturity); vendor criteria (stability, support availability, training, availability, growth direction); cost (acquisition cost, implementation cost, maintenance cost) | importance of each feature or function: ranking following the next scale: must, nice, not required; tool and vendor criteria: 5-grade scale; costs: low, medium, high |
| Prioritize Risk (rank the risks in terms of their impact on the project) | deviation of schedule from plan; deviation of effort from plan; deviation of cost from plan; likelihood of occurrence; risk exposure; risk magnitude; type: {direct, indirect}; resource: {organization, funding, people, time, business risks, technical risks, scope risks, technological risks, external dependency risks, schedule risks} | ranking according to the risk exposure; risk magnitude may be calculated in addition. |
| Prioritize Use Cases (select a certain number of scenarios and use cases to be analyzed and designed) | benefit of the scenario to the stakeholders: {critical, important, useful}; architectural impact of the scenario: {none, extends, modifies}; risks to be mitigated: {performance, availability of a product, suitability of a component}; completion of the coverage of the architecture; demonstration to the user | selection following the architectural significance: substantial architectural coverage, specific architectural point, delicate architectural point. |

Based on the information from Table 2, the strategy *by problem investigation* allows identifying the requirements for L2 decisions. A summary of these requirements is given in the table 3.

**Table 3.** Identify requirements for MC methods by problem investigation.

| Requirements for MC methods | Tools | Risks | Use cases |
|---|---|---|---|
| *Operations* | | | |
| Retain problem type | choice | ranking | choice |
| Calculate alternatives number | medium | great | great |
| Retain alternatives nature | discrete | discrete | discrete |
| Retain criteria data type | quantitative | mixed | mixed, fuzzy |
| Retain weighting type | Yes, simple | | |
| *Usage* | | | |
| Tool | | | yes |
| Easiness | easy | | |
| Skills | week | | |

For selecting MC method, we used the following process. Within the first iteration, we try to find a MC method that matches all requirements in each case.

Figure 6. illustrates the first iteration. For three considered examples, we have retained the corresponding MC method characteristics. If a MC method satisfies a given characteristic, we add "1", if does not satisfy, "0".

| Case | Tools | | | | Risks | | | | Use cases | | | |
|---|---|---|---|---|---|---|---|---|---|---|---|---|
| MC method | MAU | AHP | Out. | Wei. | MAU | AHP | Out. | Wei. | MAU | AHP | Out. | Wei. |
| Retain problem type | 1 | 1 | 1 | 1 | 1 | 1 | 1 | 1 | 1 | 1 | 1 | 1 |
| Calculate alternatives number | 1 | 0 | 1 | 1 | 1 | 0 | 1 | 1 | 1 | 0 | 1 | 1 |
| Retain alternatives nature | 1 | 1 | 1 | 1 | 1 | 1 | 1 | 1 | 1 | 1 | 1 | 1 |
| Retain criteria data type | 1 | 1 | 1 | 1 | 1 | 1 | 1 | 0 | 0 | 0 | 0 | 0 |
| Retain weighting type | 1 | 1 | 1 | 1 | | | | | | | | |
| Tool | | | | | | | | | 0 | 1 | 1 | 1 |
| Easiness | 0 | 1 | 0 | 1 | | | | | | | | |
| Skills | 0 | 0 | 0 | 1 | | | | | | | | |

1 - MC method matches requirement     - selected MC method
0 - MC method does not match requirement     - requirement is not expressed

**Fig. 6.** *MC method selection results (first iteration).*

For tools prioritization, only the weighting method satisfies all requirements. With reference to risks analysis, two MC methods are found: MAUT and outranking. To make our final choice, the engineer decides to chose methods offering a tool. So, the outranking method allowing a tool panel (PROMETHEE I and II, ELECTRE II and III (Bouyssous, 2001)) is selected. Regarding use cases prioritization, no MC method that matches requirement for criteria data type. In this case, another set of candidate methods must be considered (for example, fuzzy methods) or some requirements removed (if it is possible to remove not-satisfied requirement in the given situation).

For the lack of space, we do not consider the application of selected methods. Our aim is to illustrate, firstly, the MC method selection based on two levels requirements and, secondly, the specific situation consideration expressed by these requirements.

## 4. Conclusion

Decision-making is a difficult process and prioritizing alternatives is a good and efficient way to improve development processes. This is usually done on an intuitive way. Our aim was to offer a scientifically founded way to make this priorization by offering a guidance process to the engineer. This process proposes to use the integration of multicriteria methods to choose the most adapted alternative to each situation. We illustrated this process with examples taken within the Rational Unified Process (RUP) (Rational Rose, 2007; Kruchten, 1998). We showed how to use IP to integrate MC methods at a specific decision-making point.

Our research perspectives include improving the MC methods signatures to better select them; developing a tool that offers a systematic guidance of IP; defining MC methods as a method fragments for their integration into any existing methodologies; and exploring the issue of adapting DM methods to the situation at hand. Several extensive case studies in the IS engineering area have also been undertaken.